\RequirePackage{fixltx2e}
\documentclass[a4paper,fleqn,reqno,english,journal=cmatex,manuscript=article]{revtex4-2}
\usepackage[T1]{fontenc}
\usepackage[latin9]{inputenc}
\usepackage{color}
\usepackage{babel}
\usepackage{array}
\usepackage{float}
\usepackage{booktabs}
\usepackage{multirow}
\usepackage{amsmath}
\usepackage{amsthm}
\usepackage{amssymb}
\usepackage{graphicx}
\usepackage{wasysym}
\usepackage[unicode=true,pdfusetitle,
 bookmarks=true,bookmarksnumbered=false,bookmarksopen=false,
 breaklinks=false,pdfborder={0 0 1},backref=false,colorlinks=false]
 {hyperref}

\makeatletter

\newcommand{\lyxmathsym}[1]{\ifmmode\begingroup\def\b@ld{bold}
  \text{\ifx\math@version\b@ld\bfseries\fi#1}\endgroup\else#1\fi}

\providecommand{\tabularnewline}{\\}

\usepackage{lmodern}

\makeatother

\begin{document}
\title{Unveiling the impact of trivalent metal cation transmutation on Cs$_{2}$AgM(III)Cl$_{6}$
double perovskites using many-body perturbation theory}
\author{Surajit Adhikari{*}, Priya Johari}
\email{sa731@snu.edu.in, priya.johari@snu.edu.in}

\affiliation{Department of Physics, School of Natural Sciences, Shiv Nadar Institution
of Eminence, Greater Noida, Gautam Buddha Nagar, Uttar Pradesh 201314,
India.}
\begin{abstract}
Lead-free halide double perovskites A$_{2}$M(I)M(III)X$_{6}$ have
garnered significant attention in the past decade as promising alternatives
to CsPbX$_{3}$ perovskites, addressing concerns related to lead toxicity
and material instability. In this work, we employ a trivalent metal
cation transmutation strategy to design a series of inorganic Pb-free
halide double perovskites Cs$_{2}$AgM(III)Cl$_{6}$ and perform a
comprehensive investigation into their potential for applications
in optoelectronic devices. Our first-principles calculations, rooted
in density functional theory, demonstrate that these materials possess
a face-centered cubic lattice structure while showcasing remarkable
thermodynamic, dynamical, and mechanical stability. The G$_{0}$W$_{0}$@PBE
electronic bandgap ranges from 1.47$\lyxmathsym{\textendash}$6.20
eV, while the Bethe-Salpeter equation (BSE) indicates strong optical
absorption spanning near-infrared to ultraviolet regions for these
compounds. Furthermore, the excitonic properties suggest that these
perovskites exhibit intermediate exciton binding energies (0.17 to
0.60\,eV) and generally longer exciton lifetimes, except for the materials
with M(III) = Sc, Y, Tb, and Lu. The Fr\"ohlich model indicates that
these materials exhibit intermediate to strong carrier-phonon interactions,
with hole-phonon coupling more prominent than electron-phonon coupling.
Interestingly, the charge-separated polaronic states are found to
be less stable than the bound exciton states, with higher polaron
mobility for electrons (4.92$-$29.03 cm$^{2}$V$^{-1}$s$^{-1}$)
than for holes (0.56$-$8.69 cm$^{2}$V$^{-1}$s$^{-1}$) in these
materials. Overall, our study demonstrates that trivalent metal cation
transmutation in Cs$_{2}$AgM(III)Cl$_{6}$ enables the creation of
stable and lead-free halide double perovskites with exceptional, tunable
optoelectronic properties, making them ideal for flexible optoelectronic
applications.
\end{abstract}
\maketitle

\section{Introduction:}

Lead halide perovskites APbX$_{3}$ (A = Cs, CH$_{3}$NH$_{3}$; X
= Cl, Br, and I) have gained significant attention over the past decade
in the field of optoelectronics due to their outstanding properties$-$such
as optimal bandgaps, high carrier mobility, long diffusion lengths,
strong light absorption, low exciton binding energies, and excellent
defect tolerance \citep{chapter1-1,chapter1-2,chapter1-3,chapter1-4,chapter1-5,chapter1-6,chapter1-7,chapter1-8}.
These features make them ideal for various applications, including
high-efficiency solar cells, LEDs, lasers, sensors, and photodetectors
\citep{chapter1-9,chapter1-10,chapter1-11,chapter1-12,chapter1-13}.
Notably, their power conversion efficiency has surged from 3.8\% to
26.7\%, surpassing some Si-based thin-film technologies, establishing
them as leading low-cost materials for optoelectronic applications
\citep{chapter2-12,chapter2-53}.

Despite their excellent optoelectronic properties, organic-inorganic
lead halide perovskites face two major challenges: instability due
to the volatile CH$_{3}$NH$_{3}$$^{+}$ cation and toxicity from
Pb$^{2+}$ \citep{chapter1-16}. To overcome these, lead-free halide
perovskites are being actively explored. One approach is to replace
Pb$^{2+}$ with isovalent cations like Sn$^{2+}$ or Ge$^{2+}$, but
these tend to oxidize to the 4$+$ state, leading to material instability
\citep{chapter1-17,chapter1-18}.

To address lead toxicity, a novel approach called atomic transmutation
has emerged, where two Pb$^{2+}$ ions are replaced by one monovalent
{[}M(I){]} and one trivalent {[}M(III){]} cation, forming quaternary
double perovskites with the formula A$_{2}$M(I)M(III)X$_{6}$ (A
= monovalent cation; X = Cl, Br, or I) \citep{chapter1-19,chapter1-20}.
This strategy has led to the development of new, high-performance
materials \citep{chapter1-21,chapter1-22}.\textcolor{blue}{{} }A careful
selection of the A-site cation is essential for achieving stable and
efficient double perovskites. Moreover, the stability of these materials
is influenced by the halide anions; specifically, stability decreases
in the order of Cl- to Br- to I-containing compounds, corresponding
to the increasing ionic radius of the halide anions from Cl (1.81
$\textrm{\AA}$) to Br (1.96 $\textrm{\AA}$) to I (2.20 $\textrm{\AA}$).\textcolor{blue}{{}
}Despite these advances, the availability of suitable non-toxic monovalent
and trivalent cation combinations to replace Pb$^{2+}$ remains limited.
Notably, combinations such as A = Cs$^{+}$; M(I) = Cu$^{+}$, Ag$^{+}$,
Na$^{+}$, K$^{+}$, Rb$^{+}$; and M(III) = In$^{3+}$, Sb$^{3+}$,
Bi$^{3+}$ have led to the discovery of the \textquotedblleft elpasolite\textquotedblright{}
family of double perovskites \citep{chapter1-19,chapter1-23}.

It is well-known that the optoelectronic properties of A$_{2}$M(I)M(III)X$_{6}$
halide double perovskites (HDPs) are primarily determined by the {[}M(I)X$_{6}${]}$^{-5}$
and {[}M(III)X$_{6}${]}$^{-3}$ octahedral units. Consequently, a
strategic combination of monovalent/trivalent metal cations and halide
anions in these HDPs can lead to distinct structural features and
tunable optoelectronic behavior. This makes these double perovskite
compounds ideal model systems to explore cationic transmutation effects
for designing novel functional materials. Numerous studies have already
explored how substituting monovalent metal cations influences the
optoelectronic properties of HDP compounds \citep{chapter1-46,chapter2-38}.
In particular, Ag-based HDPs, such as Cs$_{2}$AgM(III)Cl$_{6}$,
have garnered significant attention owing to their non-toxic elemental
composition, enhanced structural stability, and tunable optoelectronic
characteristics. For example, Cs$_{2}$AgScCl$_{6}$, Cs$_{2}$AgInCl$_{6}$,
Cs$_{2}$AgSbCl$_{6}$, Cs$_{2}$AgTlCl$_{6}$, and Cs$_{2}$AgBiCl$_{6}$
HDP compounds have been experimentally synthesized, and except for
Cs$_{2}$AgScCl$_{6}$, all exhibit experimental bandgaps ranging
from 1.96 to 3.30\,eV \citep{chapter8-2,chapter1-38,chapter1-23,chapter8-6,chapter1-43}.
In addition, first-principles calculations have been employed to theoretically
predict the structural and electronic properties of several Ag-based
HDPs, including Cs$_{2}$AgAlCl$_{6}$, Cs$_{2}$AgGaCl$_{6}$, Cs$_{2}$AgAsCl$_{6}$,
and Cs$_{2}$AgRhCl$_{6}$ \citep{chapter8-1,chapter8-3,chapter8-4,chapter8-5}.
Nevertheless, comprehensive investigations into the transport, excitonic,
and polaronic properties of these HDPs are still lacking. Exciton
formation plays a pivotal role in determining charge separation efficiency
in optoelectronic devices, making the precise evaluation of excitonic
parameters (such as exciton binding energy, exciton radius, and exciton
lifetime) essential for comprehensive material characterization. At
the same time, polaron formation, which arises from electron-phonon
coupling, significantly affects exciton behavior and charge transport
in HDPs \citep{chapter2-9,chapter2-13}. Since carrier mobility is
closely linked to the strength of this coupling, a thorough understanding
of its influence on polaron mobility is crucial for optimizing the
performance of HDPs.

In this study, we employed first-principles density functional theory
(DFT) \citep{chapter2-36,chapter2-37} and many-body perturbation
theory (MBPT) \citep{chapter3-1,chapter3-2} to systematically investigate
how trivalent metal cation transmutation influences the phase stability
and optoelectronic properties of lead-free Cs$_{2}$AgM(III)Cl$_{6}$
HDPs, where M(III) spans a wide range of elements, including Al, Sc,
Ga, As, Y, Nb, Mo, Rh, In, Sb, Eu, Tb, Lu, Au, Tl, and Bi. Our calculations
reveal that all Cs$_{2}$AgM(III)Cl$_{6}$ compounds retain the conventional
cubic double perovskite structure and that most of them exhibit thermodynamic,
dynamical, and mechanical stability. Furthermore, a detailed investigation
of the electronic properties was carried out using both the HSE06
exchange-correlation (xc) functional and the G$_{0}$W$_{0}$@PBE
approach \citep{chapter1-69,chapter1-70}, revealing that these HDPs
exhibit G$_{0}$W$_{0}$@PBE bandgaps in the range of 1.47 to 6.20\,eV.
Following that, we computed the optical response by solving the Bethe-Salpeter
equation (BSE) \citep{chapter1-67,chapter1-68} on top of G$_{0}$W$_{0}$@PBE,
thereby extracting both the electronic contribution to the dielectric
function and the exciton binding energy. The results indicate that
these perovskites possess intermediate exciton binding energies and
generally longer exciton lifetimes across most HDPs. Finally, we investigated
the impact of electron-phonon coupling and estimated the polaron mobility
using the Fr\"ohlich mesoscopic model \citep{chapter2-51} and the Hellwarth
polaron model \citep{chapter2-22}, respectively. The Fr\"ohlich model
reveals intermediate to strong carrier-phonon coupling in these materials,
with hole-phonon interactions being more pronounced than those of
electrons. Overall, this work offers a comprehensive study of Cs$_{2}$AgM(III)Cl$_{6}$
compounds, highlighting their potential as lead-free perovskites with
superior optoelectronic properties for next-generation devices.

\section{Computational Details:}

All calculations were carried out using density functional theory
(DFT) \citep{chapter2-36,chapter2-37}, density functional perturbation
theory (DFPT) \citep{chapter1-60}, and many-body perturbation theory
(MBPT) \citep{chapter3-1,chapter3-2} as implemented in the Vienna
Ab initio Simulation Package (VASP) \citep{chapter1-31,chapter1-32}.
The electron-ion interactions were modeled using the projector augmented
wave (PAW) pseudopotential method \citep{chapter1-33}. The structural
optimizations were performed using the Perdew-Burke-Ernzerhof (PBE)
xc functional within the framework of the generalized gradient approximation
(GGA) \citep{chapter1-34}, which incorporates electron-electron interactions.
The structural optimization of the systems under consideration was
carried out using a $4\times4\times4$ $\Gamma$-centered $\mathrm{k}$-point
grid. A plane-wave cutoff energy of 400 eV was applied to expand the
wave function with periodic boundary conditions imposed in all directions.
The unit cell dimensions and ionic positions were optimized until
the forces on each ion and the total energy reached convergence with
tolerances of less than 0.01 eV/$\textrm{\AA}$ and 10$^{-6}$ eV, respectively.
To assess the dynamical stability of these compounds, we computed
the phonon dispersion curves employing DFPT with $2\times2\times2$
supercells, utilizing the PHONOPY package \citep{chapter3-6}. Initially,
the electronic band structures were calculated using the PBE xc functional,
with the inclusion of spin-orbit coupling (SOC) effects. However,
the PBE functional tends to underestimate the energy bandgaps due
to the self-interaction error, particularly for these HDPs. To obtain
a more accurate estimation, the electronic bandgaps were calculated
using both the hybrid HSE06 (exact exchange = 25\%) xc functional
\citep{chapter1-35} and the MBPT-based G$_{0}$W$_{0}$@PBE method
\citep{chapter1-69,chapter1-70}. The effective masses of the charge
carriers were determined using the SUMO code \citep{chapter2-10},
which applied a parabolic fit at the band edges. Additionally, the
density of states (DOS) calculations were performed employing the
hybrid HSE06 functional with a $\Gamma$-centered $6\times6\times6$
$\mathrm{k}$-point grid. In addition, the optical properties were
determined by solving the Bethe-Salpeter equation (BSE) \citep{chapter1-67,chapter1-68}
on top of the G$_{0}$W$_{0}$@PBE calculations, which account for
electron-hole interactions. For the GW-BSE calculations, a $\Gamma$-centered
$4\times4\times4$ $\mathrm{k}$-grid and 540 converged NBANDS were
utilized. The electron-hole kernel for the BSE was constructed using
6 occupied and 6 unoccupied bands. Post-processing of the elastic
and optical properties was carried out with the VASPKIT package \citep{chapter1-48}.
Additionally, the ionic contribution to the dielectric constant was
computed using the DFPT method.

\section{Results and Discussions:}

To evaluate the impact of trivalent metal cation transmutation on
the optoelectronic properties of halide double perovskites (HDPs),
we performed a comprehensive study of Cs$_{2}$AgM(III)Cl$_{6}$,
where M(III) includes Al, Sc, Ga, As, Y, Nb, Mo, Rh, In, Sb, Eu, Tb,
Lu, Au, Tl, and Bi. The stability of these HDPs, along with their
structural, electronic, optical, excitonic, and polaronic properties
are thoroughly analyzed and discussed in the following sections. This
investigation aims to provide theoretical insights and offer guidance
for future experimental research.

\subsection{Structural Properties:}

\begin{figure}[H]
\begin{centering}
\includegraphics[width=1\textwidth,height=1\textheight,keepaspectratio]{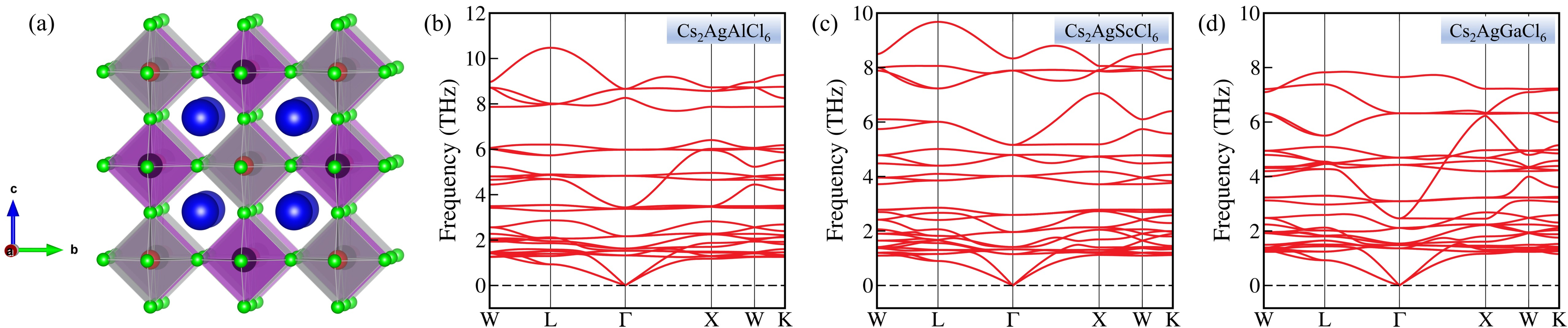}
\par\end{centering}
\caption{\label{Figure:1}(a) Polyhedral view of the Cs$_{2}$AgM(III)Cl$_{6}$
HDPs (space group $Fm\bar{3}m$), where blue, red, black, and green
balls represent Cs, Ag, M(III), and Cl atoms, respectively. Also,
phonon dispersion curves of (b) Cs$_{2}$AgAlCl$_{6}$, (c) Cs$_{2}$AgScCl$_{6}$,
and (d) Cs$_{2}$AgGaCl$_{6}$ HDP, calculated with the DFPT method.}

\end{figure}

\subsubsection{Crystal Structure and Crystallographic Stability:}

The crystal structure of halide double perovskites (HDPs) is a key
factor influencing their optoelectronic properties. \ref{Figure:1}(a)
illustrates the crystal structure of Cs$_{2}$AgM(III)Cl$_{6}$ HDPs,
where M(III) represents trivalent metal cations, including Al$^{3+}$,
Sc$^{3+}$, Ga$^{3+}$, As$^{3+}$, Y$^{3+}$, Nb$^{3+}$, Mo$^{3+}$,
Rh$^{3+}$, In$^{3+}$, Sb$^{3+}$, Eu$^{3+}$, Tb$^{3+}$, Lu$^{3+}$,
Au$^{3+}$, Tl$^{3+}$, and Bi$^{3+}$. The crystal structure calculations
are carried out using a conventional unit cell of Cs$_{2}$AgM(III)Cl$_{6}$,
which consists of 40 atoms. Our findings indicate that, under ambient
conditions, all the HDPs exhibit a typical face-centered cubic lattice
with the space group $Fm\bar{3}m$. In this configuration, Cs atoms
occupy the Wyckoff position 8c (0.25, 0.25, 0.25), while the Ag and
M(III) atoms are located at the Wyckoff positions 4b (0.5, 0.5, 0.5)
and 4a (0, 0, 0), respectively. The Cl atoms are positioned at the
Wyckoff site 24e (x, 0, 0). The structure of Cs$_{2}$AgM(III)Cl$_{6}$
is built from alternating $[\mathrm{AgCl_{6}}]^{-5}$ and $[\mathrm{M(III)X_{6}}]^{-3}$
octahedra that are regularly connected at their corners to create
a three-dimensional framework. The Cs$^{+}$ ions are situated at
the centers of the cavities within this framework. The optimized lattice
parameters and bond lengths {[}Ag-Cl and M(III)-Cl{]} for the examined
HDPs, calculated using the PBE xc functional, are presented in Table
\ref{Table:1} and Table S1, respectively. The computed lattice parameters
are in excellent agreement with both previous theoretical and experimental
studies \citep{chapter1-23,chapter1-38,chapter1-43,chapter1-46,chapter8-1,chapter8-2,chapter8-3,chapter8-4,chapter8-5,chapter8-6,chapter8-7}.

\begin{table}[H]
\caption{\label{Table:1}Optimized lattice parameters (in unit $\textrm{\AA}$),
Goldschmidt\textquoteright s tolerance factor ($t$), octahedral factor
($\mu$), new tolerance factor ($\tau$), and the stability of Cs$_{2}$AgM(III)Cl$_{6}$
HDPs are listed. Here, $t$ and $e$ represent theoretical and experimental
lattice parameters, respectively.}

\centering{}%
\begin{tabular}{ccccccccc}
\toprule 
\multirow{2}{*}{{\footnotesize{}Compounds}} & \multicolumn{2}{c}{{\footnotesize{}Lattice parameters (a = b = c), $\textrm{\AA}$}} & \multirow{2}{*}{{\footnotesize{}$t$}} & \multirow{2}{*}{{\footnotesize{}$\mu$}} & \multirow{2}{*}{{\footnotesize{}$\tau$}} & \multicolumn{3}{c}{{\footnotesize{}Stability}}\tabularnewline
\cmidrule{2-3} \cmidrule{3-3} \cmidrule{7-9} \cmidrule{8-9} \cmidrule{9-9} 
 & {\footnotesize{}In this study} & {\footnotesize{}Previous report} &  &  &  & {\footnotesize{}Thermodynamical} & {\footnotesize{}Dynamical} & {\footnotesize{}Mechanical}\tabularnewline
\midrule
{\footnotesize{}Cs$_{2}$AgAlCl$_{6}$} & {\footnotesize{}10.36} & {\footnotesize{}10.11$_{t}$ \citep{chapter8-1}} & {\footnotesize{}0.98} & {\footnotesize{}0.47} & {\footnotesize{}3.93} & {\footnotesize{}$\checked$} & {\footnotesize{}$\checked$} & {\footnotesize{}$\checked$}\tabularnewline
{\footnotesize{}Cs$_{2}$AgScCl$_{6}$} & {\footnotesize{}10.59} & {\footnotesize{}10.47$_{e}$ \citep{chapter8-2}} & {\footnotesize{}0.95} & {\footnotesize{}0.52} & {\footnotesize{}3.81} & {\footnotesize{}$\checked$} & {\footnotesize{}$\checked$} & {\footnotesize{}$\checked$}\tabularnewline
{\footnotesize{}Cs$_{2}$AgGaCl$_{6}$} & {\footnotesize{}10.41} & {\footnotesize{}10.32$_{t}$ \citep{chapter8-3}} & {\footnotesize{}0.97} & {\footnotesize{}0.49} & {\footnotesize{}3.86} & {\footnotesize{}$\checked$} & {\footnotesize{}$\checked$} & {\footnotesize{}$\checked$}\tabularnewline
{\footnotesize{}Cs$_{2}$AgAsCl$_{6}$} & {\footnotesize{}10.57} & {\footnotesize{}10.52$_{t}$ \citep{chapter8-4}} & {\footnotesize{}0.98} & {\footnotesize{}0.48} & {\footnotesize{}3.89} & {\footnotesize{}$\checked$} & {\footnotesize{}$\times$} & {\footnotesize{}$\checked$}\tabularnewline
{\footnotesize{}Cs$_{2}$AgYCl$_{6}$} & {\footnotesize{}10.84} &  & {\footnotesize{}0.92} & {\footnotesize{}0.57} & {\footnotesize{}3.79} & {\footnotesize{}$\checked$} & {\footnotesize{}$\checked$} & {\footnotesize{}$\checked$}\tabularnewline
{\footnotesize{}Cs$_{2}$AgNbCl$_{6}$} & {\footnotesize{}10.53} &  & {\footnotesize{}0.95} & {\footnotesize{}0.52} & {\footnotesize{}3.81} & {\footnotesize{}$\times$} & {\footnotesize{}$\checked$} & {\footnotesize{}$\checked$}\tabularnewline
{\footnotesize{}Cs$_{2}$AgMoCl$_{6}$} & {\footnotesize{}10.44} &  & {\footnotesize{}0.96} & {\footnotesize{}0.51} & {\footnotesize{}3.83} & {\footnotesize{}$\times$} & {\footnotesize{}$\checked$} & {\footnotesize{}$\checked$}\tabularnewline
{\footnotesize{}Cs$_{2}$AgRhCl$_{6}$} & {\footnotesize{}10.29} & {\footnotesize{}10.09$_{t}$ \citep{chapter8-5}} & {\footnotesize{}0.96} & {\footnotesize{}0.50} & {\footnotesize{}3.84} & {\footnotesize{}$\checked$} & {\footnotesize{}$\checked$} & {\footnotesize{}$\checked$}\tabularnewline
{\footnotesize{}Cs$_{2}$AgInCl$_{6}$} & {\footnotesize{}10.65} & {\footnotesize{}10.48$_{e}$ \citep{chapter1-38}, 10.68$_{t}$ \citep{chapter8-7}} & {\footnotesize{}0.94} & {\footnotesize{}0.54} & {\footnotesize{}3.79} & {\footnotesize{}$\checked$} & {\footnotesize{}$\checked$} & {\footnotesize{}$\checked$}\tabularnewline
{\footnotesize{}Cs$_{2}$AgSbCl$_{6}$} & {\footnotesize{}10.84} & {\footnotesize{}10.70$_{e}$ \citep{chapter1-23}, 10.56$_{t}$ \citep{chapter1-23}} & {\footnotesize{}0.94} & {\footnotesize{}0.53} & {\footnotesize{}3.80} & {\footnotesize{}$\checked$} & {\footnotesize{}$\checked$} & {\footnotesize{}$\checked$}\tabularnewline
{\footnotesize{}Cs$_{2}$AgEuCl$_{6}$} & {\footnotesize{}10.81} &  & {\footnotesize{}0.91} & {\footnotesize{}0.58} & {\footnotesize{}3.80} & {\footnotesize{}$\checked$} & {\footnotesize{}$\times$} & {\footnotesize{}$\checked$}\tabularnewline
{\footnotesize{}Cs$_{2}$AgTbCl$_{6}$} & {\footnotesize{}10.86} &  & {\footnotesize{}0.92} & {\footnotesize{}0.57} & {\footnotesize{}3.79} & {\footnotesize{}$\checked$} & {\footnotesize{}$\checked$} & {\footnotesize{}$\checked$}\tabularnewline
{\footnotesize{}Cs$_{2}$AgLuCl$_{6}$} & {\footnotesize{}10.75} &  & {\footnotesize{}0.93} & {\footnotesize{}0.56} & {\footnotesize{}3.79} & {\footnotesize{}$\checked$} & {\footnotesize{}$\checked$} & {\footnotesize{}$\checked$}\tabularnewline
{\footnotesize{}Cs$_{2}$AgAuCl$_{6}$} & {\footnotesize{}10.40} &  & {\footnotesize{}0.93} & {\footnotesize{}0.55} & {\footnotesize{}3.79} & {\footnotesize{}$\checked$} & {\footnotesize{}$\times$} & {\footnotesize{}$\checked$}\tabularnewline
{\footnotesize{}Cs$_{2}$AgTlCl$_{6}$} & {\footnotesize{}10.74} & {\footnotesize{}10.56$_{e}$ \citep{chapter8-6}} & {\footnotesize{}0.92} & {\footnotesize{}0.56} & {\footnotesize{}3.79} & {\footnotesize{}$\checked$} & {\footnotesize{}$\times$} & {\footnotesize{}$\checked$}\tabularnewline
{\footnotesize{}Cs$_{2}$AgBiCl$_{6}$} & {\footnotesize{}10.94} & {\footnotesize{}10.78$_{e}$ \citep{chapter1-43}, 10.94$_{t}$ \citep{chapter1-46}} & {\footnotesize{}0.90} & {\footnotesize{}0.60} & {\footnotesize{}3.82} & {\footnotesize{}$\checked$} & {\footnotesize{}$\times$} & {\footnotesize{}$\checked$}\tabularnewline
\bottomrule
\end{tabular}
\end{table}

We now turn our attention to the crystallographic stability of the
HDPs under consideration. The general crystallographic framework of
these materials follows the stoichiometry A$_{2}$M(I)M(III)X$_{6}$.
In this study, the A-site is occupied by Cs, the monovalent metal
cation {[}M(I){]} is Ag, and the trivalent metal cation {[}M(III){]}
is selected as Al, Sc, Ga, As, Y, Nb, Mo, Rh, In, Sb, Eu, Tb, Lu,
Au, Tl, and Bi, while X is the halide component Cl. Typically, the
crystallographic stability of perovskites is quantitatively evaluated
using Goldschmidt\textquoteright s empirical criteria, which rely
on two key parameters: the Goldschmidt tolerance factor ($t$) and
the octahedral factor ($\mu$) \citep{chapter1-36,chapter1-37,chapter1-38}.
Recently, an additional parameter, referred to as the new tolerance
factor ($\tau$), has also been introduced \citep{chapter1-39}. The
relationships governing these three parameters are expressed by the
following equations:
\begin{center}
\begin{equation}
t=\frac{(r_{A}+r_{X})}{\sqrt{2}\left[\frac{\left(r_{M(I)}+r_{M(III)}\right)}{2}+r_{X}\right]},
\end{equation}
\par\end{center}

\begin{center}
\begin{equation}
\mu=\frac{\left(r_{M(I)}+r_{M(III)}\right)}{2r_{X}},
\end{equation}
\par\end{center}

\begin{center}
\begin{equation}
\tau=\frac{2r_{X}}{\left(r_{M(I)}+r_{M(III)}\right)}-n_{A}\left(n_{A}-\frac{2r_{A}/\left(r_{M(I)}+r_{M(III)}\right)}{ln\left[2r_{A}/\left(r_{M(I)}+r_{M(III)}\right)\right]}\right).
\end{equation}
\par\end{center}

The symbols $r_{A}$, $r_{M(I)}$, $r_{M(III)}$, and $r_{X}$ represent
the ionic radii of the respective ions, whereas $n_{A}$ denotes the
oxidation state of the cation occupying the A-site. Based on previous
studies, cubic perovskite structures are typically stable when the
values of $\mu$ and $t$ fall within the ranges $0.442<\mu<0.895$
and $0.80<t<1.10$, respectively \citep{chapter1-37,chapter2-2}.
Additionally, $\tau<4.18$ indicates structural stability for perovskites
\citep{chapter1-39}. Table \ref{Table:1} indicates that the considered
HDPs fulfill all crystallographic stability criteria, providing strong
evidence for the formation of a stable cubic phase in Cs$_{2}$AgM(III)Cl$_{6}$
HDPs.

\subsubsection{Thermodynamic Stabilty:}

In addition to the crystallographic stability, the decomposition energy
of Cs$_{2}$AgM(III)Cl$_{6}$ HDPs is also calculated using the PBE
xc functional to determine their thermodynamic stability. Since HDP
contains multiple cations, it exhibits a diverse range of decomposition
pathways. The most straightforward decomposition route for these HDP
materials involves their breakdown into the corresponding binary products,
as the majority of these compounds are synthesized via the reverse
reaction \citep{chapter1-46,chapter2-38}. Additionally, the enthalpy
of decomposition can be evaluated for both binary and ternary products.
The decomposition enthalpy ($\Delta\mathrm{H_{D}}$) is defined as
the difference between the total energy of the decomposition products
and that of the host material (for details, see the Supplemental Material).
Accordingly, the expression for $\Delta\mathrm{H_{D}}$ corresponding
to the two distinct decomposition pathways can be written as follows
\citep{chapter1-21,chapter1-46}:

Pathway (1)

\begin{equation}
\mathrm{\Delta H_{D_{1}}=2E_{T}[CsCl]+E_{T}[AgCl]+E_{T}[M(III)Cl_{3}]-E_{T}[Cs_{2}AgM(III)Cl_{6}]}\label{eq:4}
\end{equation}

Pathway (2)

\begin{equation}
\mathrm{\Delta H_{D_{2}}=1/2E_{T}[CsCl]+E_{T}[AgCl]+1/2E_{T}[Cs_{3}M_{2}(III)Cl_{9}]-E_{T}[Cs_{2}AgM(III)Cl_{6}]}\label{eq:5}
\end{equation}

where $\mathrm{E_{T}}$ represents the total energy of the corresponding
compounds.\textcolor{red}{{} }Our computational results for the two
decomposition pathways discussed above are summarized in Table S2.
From equation \ref{eq:4}, it is evident that all the HDPs studied
exhibit positive decomposition energies ($\geq$ 30 meV/atom), except
those based on M(III) = Nb and Mo. The positive value of $\Delta\mathrm{H_{D}}$
signifies the energy acquired from the binary compositions by the
host material, emphasizing the thermodynamic stability of the host
material. On the other hand, when the ternary decomposition pathway
(equation \ref{eq:5}) is considered, the decomposition enthalpy generally
decreases for all HDPs, with some exhibiting even more negative values.
This indicates that the stability of the HDP is compromised, making
it prone to decomposition into ternary compounds. Consequently, theoretical
studies must consider all potential decomposition pathways, including
both binary and ternary compounds, to accurately assess the stability
of the HDP. Overall, our findings show that the majority of the HDPs
examined exhibit excellent stability against decomposition, with the
exception of Nb- and Mo-based compounds.

\subsubsection{Dynamical Stabilty:}

Subsequently, the dynamical stability of Cs$_{2}$AgM(III)Cl$_{6}$
HDPs is assessed, as it plays a pivotal role in material stability
and is intrinsically linked to the characteristics of phonon modes.
Therefore, self-consistent phonon calculations are performed for these
16 investigated HDPs using the DFPT method \citep{chapter1-60}. The
phonon dispersion curves of three examined HDPs are shown in Figure
\ref{Figure:1}{[}(b)-(d){]}, while those for the other eight investigated
HDPs are displayed in Figure S1. For Cs$_{2}$AgM(III)Cl$_{6}$ HDPs,
the structural symmetry reveals 30 phonon modes corresponding to 10
atoms. Among these 30 phonon modes, only 3 are acoustic, while the
remaining 27 are optical, categorized into low- and high-frequency
phonons, respectively. Overall, the phonon band structures of these
eleven HDPs exhibit the absence of imaginary frequencies, indicating
the dynamical stability of these compounds at T = 0 K. In contrast,
the remaining five Cs$_{2}$AgM(III)Cl$_{6}$ HDPs, where M(III) =
As, Eu, Au, Tl, and Bi, are found to be dynamically unstable at T
= 0 K. However, among the five dynamically unstable HDPs, those based
on Tl and Bi have already been experimentally synthesized \citep{chapter8-6,chapter1-43},
suggesting that these compounds might be stable at elevated temperatures.
The study by Zhao et al. has already demonstrated that that Cs$_{2}$AgBiCl$_{6}$
remains dynamically stable at a temperature of 300 K \citep{chapter1-46}.

\subsubsection{Mechanical Stability and Elastic Properties:}

To qualitatively assess the impact of trivalent metal cation transmutation,
the mechanical stability and elastic properties of Cs$_{2}$AgM(III)Cl$_{6}$
HDPs are also explored along with their thermodynamic and dynamical
stability. The second-order elastic constants ($C_{ij}$) are computed
using the energy-strain method to evaluate the mechanical stability
\citep{chapter1-47}. For cubic symmetry, the mechanical stability
and related properties of the crystal can be fully explained using
three independent elastic constants: $C_{11}$, $C_{12}$, and $C_{44}$.
The Born stability criterion for ensuring the mechanical stability
of a cubic crystal structure is expressed as follows \citep{chapter1-47}:

\begin{equation}
C_{11}>0,\;C_{11}-C_{12}>0,\;C_{11}+2C_{12}>0,\;C_{44}>0
\end{equation}

The $C_{ij}$ values for all HDPs, shown in Table S3, satisfy the
Born stability conditions, suggesting that they exhibit excellent
mechanical stability in the cubic phase. The bulk modulus ($B$),
shear modulus ($G$), Young\textquoteright s modulus ($Y$), and Poisson\textquoteright s
ratio ($\nu$) are calculated using the Voigt$-$Reuss$-$Hill method
\citep{chapter1-49,chapter1-50} based on the given elastic constants
and are tabulated in Table S3. The fragility of these perovskites
is also studied using Pugh's ($B/G$) ratio and Poisson's ($\nu$)
ratio \citep{chapter1-51}. The calculated values of $B/G$ ($>$
1.75) and $\nu$ ($>$ 0.26) indicate that the studied HDPs are ductile
in nature (see Table S3). Also, the elastic anisotropy ($A$) of these
materials is computed using the relation: $A=2C_{44}/(C_{11}-C_{12})$
\citep{chapter1-52}. The value of $A$ is 1 for an isotropic crystal,
and any deviation from this value indicates the extent of elastic
anisotropy in the crystal. Based on the calculated results, all the
HDPs examined exhibit anisotropic behavior (see Table S3).

\begin{table}[H]
\caption{\label{Table:2}Computed bandgap ($E_{g}$) of Cs$_{2}$AgM(III)Cl$_{6}$
HDPs using the PBE/PBE+SOC, HSE06, and G$_{0}$W$_{0}$@PBE methods,
where M(III) = Al, Sc, Ga, As, Y, Rh, In, Sb, Tb, Lu, Tl, and Bi.
Here, $d$, $i$, $t$, and $e$ represent direct, indirect, theoretical,
and experimental bandgaps, respectively.}

\centering{}%
\begin{tabular}{cccccc}
\hline 
\multirow{1}{*}{Compounds} & PBE/PBE+SOC & HSE06 & G$_{0}$W$_{0}$@PBE & Nature & Previous work\tabularnewline
\hline 
Cs$_{2}$AgAlCl$_{6}$ & 2.22/2.19 & 3.75 & 4.88 & $d$ & 3.43$_{t}$ \citep{chapter8-1}\tabularnewline
Cs$_{2}$AgScCl$_{6}$ & 3.30/3.26 & 4.94 & 6.20 & $i$ & \tabularnewline
Cs$_{2}$AgGaCl$_{6}$ & 0.96/0.94 & 2.34 & 3.45 & $d$ & 1.20$_{t}$ \citep{chapter8-3}\tabularnewline
Cs$_{2}$AgAsCl$_{6}$ & 1.30/1.28 & 2.27 & 2.36 & $i$ & 2.49$_{t}$ \citep{chapter8-4}\tabularnewline
Cs$_{2}$AgYCl$_{6}$ & 3.70/3.66 & 5.32 & 5.92 & $i$ & \tabularnewline
Cs$_{2}$AgRhCl$_{6}$ & 0.58/0.42 & 2.29 & 2.31 & $d$ & 0.55$_{t}$ \citep{chapter8-5}\tabularnewline
Cs$_{2}$AgInCl$_{6}$ & 1.03/1.01 & 2.41 & 3.39 & $d$ & 3.30$_{e}$ \citep{chapter1-38}\tabularnewline
Cs$_{2}$AgSbCl$_{6}$ & 1.38/1.35 & 2.34 & 2.57 & $i$ & 2.60$_{e}$ \citep{chapter1-23}\tabularnewline
Cs$_{2}$AgTbCl$_{6}$ & 3.63/3.57 & 5.24 & 5.77 & $i$ & \tabularnewline
Cs$_{2}$AgLuCl$_{6}$ & 3.51/3.46 & 5.16 & 5.74 & $i$ & \tabularnewline
Cs$_{2}$AgTlCl$_{6}$ & 0.04/0.04 & 0.69 & 1.47 & $d$ & 1.96$_{e}$ \citep{chapter8-6}\tabularnewline
Cs$_{2}$AgBiCl$_{6}$ & 1.84/1.58 & 2.94 & 3.21 & $i$ & 2.77$_{e}$ \citep{chapter1-43}\tabularnewline
\hline 
\end{tabular}
\end{table}

\subsection{Electronic Properties:}

Once stability is confirmed, electronic structure calculations for
Cs$_{2}$AgM(III)Cl$_{6}$ HDPs are performed, as they are crucial
for designing optoelectronic devices. In this analysis, the partial
density of states (PDOS) and total density of states (TDOS) are examined
alongside the band edge positions, specifically the conduction band
minimum (CBM) and valence band maximum (VBM). Additionally, the nature
of bandgap, whether direct or indirect, is assessed to obtain a comprehensive
understanding of the electronic structure.

\begin{figure}[H]
\begin{centering}
\includegraphics[width=1\textwidth,height=1\textheight,keepaspectratio]{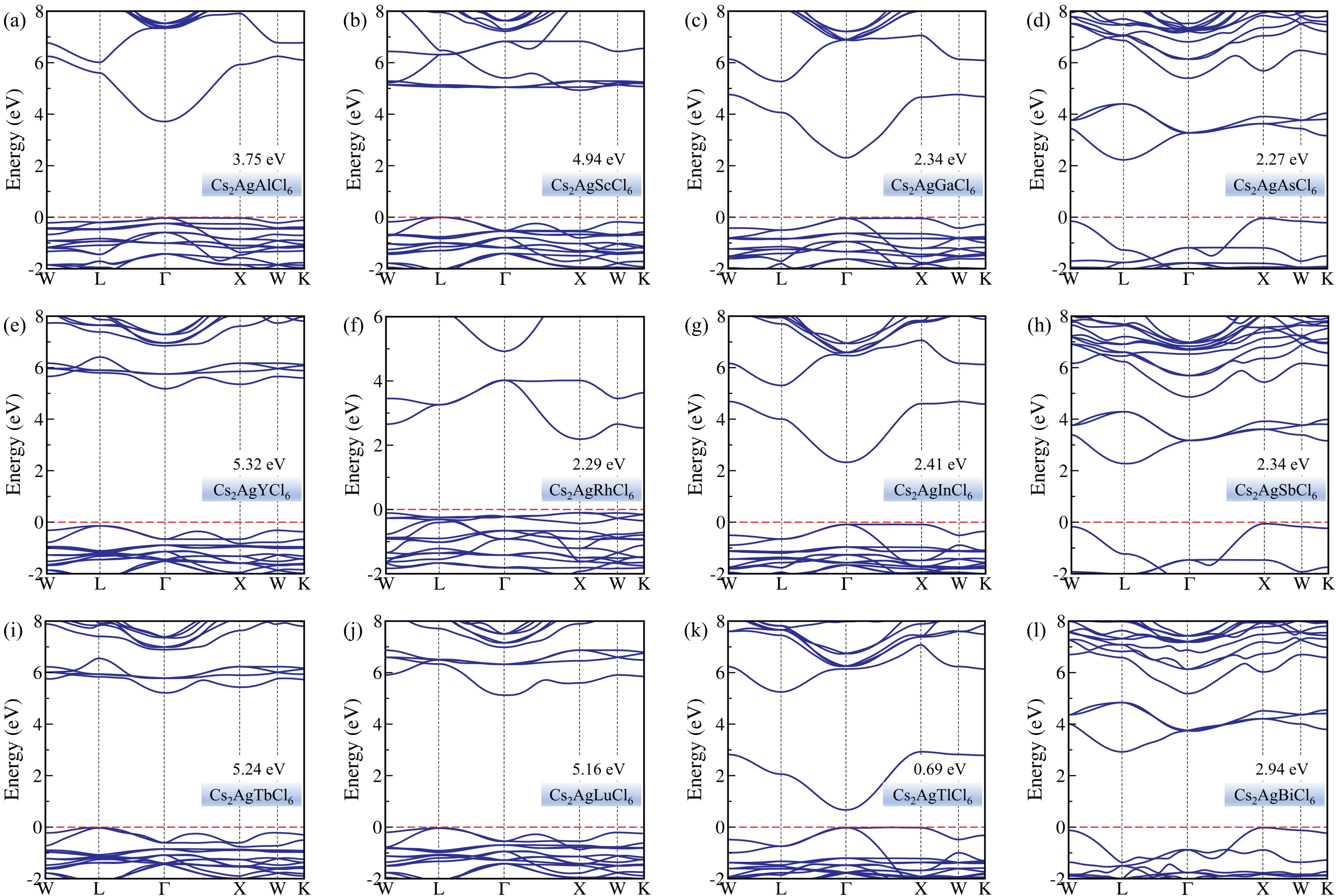}
\par\end{centering}
\caption{\label{Figure:2}Electronic band structures of Cs$_{2}$AgM(III)Cl$_{6}$
HDPs obtained using HSE06 xc functional, where M(III) = Al, Sc, Ga,
As, Y, Rh, In, Sb, Tb, Lu, Tl, and Bi. Band structures of considered
HDPs are obtained on the following path (in crystal coordinates):
W (0.5, 0.25, 0.75) - L (0.5, 0.5, 0.5) - (0, 0, 0) - X (0.5, 0, 0.5)
- W (0.5, 0.25, 0.75) - K (0.375, 0.375, 0.75). The Fermi level is
set to be zero and marked by the red line.}

\end{figure}

Initially, electronic structure calculations for investgated HDPs
are carried out using the widely used semilocal PBE xc functional,
both with and without the inclusion of spin-orbit coupling (SOC) effects.
However, PBE xc functional is widely recognized for its tendency to
underestimate the bandgap of HDPs due to the self-interaction error
\citep{chapter2-38}, a behavior that is also observed in our investigation.
In addition, it is observed that SOC does not significantly affect
the bandgap of the investigated HDPs, with the exception of Cs$_{2}$AgRhCl$_{6}$
and Cs$_{2}$AgBiCl$_{6}$ (see \ref{Table:2}). It is crucial to
incorporate SOC for bandgap calculations in these two HDPs, especially
for Cs$_{2}$AgBiCl$_{6}$ due to the heavy Bi element. To achieve
a more accurate bandgap estimation, hybrid HSE06 xc functional \citep{chapter1-35}
and the MBPT-based G$_{0}$W$_{0}$@PBE \citep{chapter1-69,chapter1-70}
method are also applied. The band structures of these Cs$_{2}$AgM(III)Cl$_{6}$
HDPs calculated using HSE06 are shown in Figure \ref{Figure:2}. The
band structure analysis revealed that HDPs with M(III) = Al, Ga, Rh,
In, and Tl display direct bandgaps, while those with M(III) = Sc,
As, Y, Sb, Tb, Lu, and Bi show indirect bandgaps. In contrast, HDPs
with M(III) = Nb, Mo, Eu, and Au exhibit metallic characteristics.
The bandgaps of these HDPs, calculated using PBE/PBE+SOC, HSE06, and
G$_{0}$W$_{0}$@PBE, are presented in Table \ref{Table:2}. In direct
bandgap materials, the CBM and VBM are situated at the $\Gamma$ point
for HDPs with M(III) elements such as Al, Ga, In, and Tl. However,
for HDPs where M(III) is Rh, both the CBM and VBM are positioned at
the X point. Conversely, for indirect bandgap materials such as Cs$_{2}$AgScCl$_{6}$,
the CBM is situated at the X point, while the VBM is found at the
L point. For HDPs incorporating pnictogen elements like M(III) = As,
Sb, and Bi, the CBM and VBM are located at the L and X points, respectively.
In contrast, for M(III) = Y, Tb, and Lu, the CBM and VBM are positioned
at the $\Gamma$ and L points, respectively. The bandgaps of these
HDPs, calculated using HSE06 and G$_{0}$W$_{0}$@PBE, are observed
to range between 0.69$\lyxmathsym{\textendash}$5.32 eV and 1.47$\lyxmathsym{\textendash}$6.20
eV, respectively. Our analysis reveals that the G$_{0}$W$_{0}$@PBE
bandgaps for these HDPs are in close agreement with the available
experimental results \citep{chapter1-23,chapter1-38,chapter1-43,chapter8-6}.
Based on this study, we can conclude that G$_{0}$W$_{0}$@PBE delivers
more precise bandgaps for these materials. However, for Cs$_{2}$AgBiCl$_{6}$,
G$_{0}$W$_{0}$@PBE+SOC yields a more accurate bandgap. The study
demonstrates that the transmutation of trivalent metal cations in
Cs$_{2}$AgM(III)Cl$_{6}$ makes it suitable for a broad spectrum
of optoelectronic devices, due to its tunable bandgap, which spans
from the visible to ultraviolet region.

To gain a deeper understanding of the electronic band structures of
these HDPs, both total and partial density of states were calculated
using the HSE06 xc functional. The corresponding results are presented
in Figure S2 of the Supplemental Material. The contributions of different
orbitals to the CBM and VBM can be seen here, which account for the
variations in band structure and bandgap among the compounds. To aid
understanding, we have summarized these orbital contributions for
each compound in Table S5.

To further understand charge carrier transport, the effective masses
of electrons ($m_{e}^{*}$) and holes ($m_{h}^{*}$) for all the HDPs
are also calculated using G$_{0}$W$_{0}$@PBE band structures. The
effective masses are determined by fitting the E$-$k dispersion curves
with the formula, $\text{\ensuremath{m^{*}=\hbar^{2}\left[\partial^{2}E(k)/\partial k^{2}\right]^{-1}}}$
and are presented in Table \ref{Table:3}. In our study, the reduced
energy dispersion of the VBM leads to higher hole effective masses
($m_{h}^{*}$) of these HDPs compared to those of electrons ($m_{e}^{*}$).
The values of $m_{e}^{*}$ and $m_{h}^{*}$ of these HDPs are in the
range of 0.267$-$0.573 and 0.559$-$2.451, respectively, indicating
that the electron mobility of these HDPs is higher than their hole
mobility. Additionally, the variation in charge-carrier mobilities
is a key factor in reducing the electron-hole recombination rate.
This difference can be quantified by the ratio of the absolute values
of the effective masses \citep{chapter8-8}, $D=\left|m_{h}^{*}\right|/\left|m_{e}^{*}\right|.$
A larger relative ratio of the effective masses ($D$) suggests a
significant disparity between electron and hole mobility, which in
turn decreases the recombination rate of electron-hole pairs. Table
\ref{Table:3} shows that all the HDPs exhibit higher values of $D$
(ranging from 1.44 to 5.43), indicating reduced electron-hole recombination
rates and consequently improving the optoelectronic efficiency of
these HDPs.

\begin{table}[H]
\caption{\label{Table:3}Computed carrier's effective mass, reduced mass, effective
mass ratio, and dielectric constant of Cs$_{2}$AgM(III)Cl$_{6}$
HDPs, where M(III) = Al, Sc, Ga, As, Y, Rh, In, Sb, Tb, Lu, Tl, and
Bi. All values of the effective mass are in terms of free-electron
mass ($m_{0}$).}

\centering{}%
\begin{tabular}{cccccccc}
\hline 
Compounds & $m_{e}^{*}$ ($m_{0}$) & $m_{h}$ ($m_{0}$) & $\mu^{*}$ ($m_{0}$) & $D$ & $\varepsilon_{\infty}$ & $\varepsilon_{ion}$ & $\varepsilon_{static}$\tabularnewline
\hline 
Cs$_{2}$AgAlCl$_{6}$ & 0.319 & 0.958 & 0.239 & 3.00 & 2.86 & 9.23 & 12.09\tabularnewline
Cs$_{2}$AgScCl$_{6}$ & 0.573 & 1.227 & 0.391 & 2.14 & 2.84 & 8.15 & 10.99\tabularnewline
Cs$_{2}$AgGaCl$_{6}$ & 0.267 & 0.559 & 0.181 & 2.09 & 3.19 & 9.66 & 12.85\tabularnewline
Cs$_{2}$AgAsCl$_{6}$ & 0.349 & 0.584 & 0.218 & 1.67 & 4.13 & 13.62 & 17.75\tabularnewline
Cs$_{2}$AgYCl$_{6}$ & 0.404 & 1.178 & 0.301 & 2.92 & 2.58 & 6.04 & 8.62\tabularnewline
Cs$_{2}$AgRhCl$_{6}$ & 0.451 & 2.451 & 0.381 & 5.43 & 4.09 & 4.45 & 8.54\tabularnewline
Cs$_{2}$AgInCl$_{6}$ & 0.275 & 0.705 & 0.198 & 2.56 & 3.11 & 7.51 & 10.62\tabularnewline
Cs$_{2}$AgSbCl$_{6}$ & 0.363 & 0.580 & 0. 223 & 1.60 & 3.77 & 10.63 & 14.40\tabularnewline
Cs$_{2}$AgTbCl$_{6}$ & 0.382 & 1.113 & 0.284 & 2.91 & 2.60 & 9.44 & 12.04\tabularnewline
Cs$_{2}$AgLuCl$_{6}$ & 0.365 & 1.284 & 0.284 & 3.52 & 2.74 & 6.24 & 8.98\tabularnewline
Cs$_{2}$AgTlCl$_{6}$ & 0.376 & 2.027 & 0.317 & 5.39 & 4.35 & 6.96 & 11.31\tabularnewline
Cs$_{2}$AgBiCl$_{6}$ & 0.389 & 0.562 & 0.230 & 1.44 & 3.64 & 10.52 & 14.16\tabularnewline
\hline 
\end{tabular}
\end{table}

\subsection{Optical Properties:}

In addition to the electronic properties discussed above, the optical
absorption features play a vital role in optoelectronic applications
and deserve detailed exploration. The optical absorption spectra are
derived from the imaginary part {[}Im($\varepsilon$){]} of the macroscopic
dielectric function, enabling us to analyze the influence of electron-hole
interactions (bound exciton states) on the absorption characteristics.
The optical response of these HDPs is determined by solving the Bethe-Salpeter
equation (BSE) on top of the G$_{0}$W$_{0}$@PBE calculations, which
explicitly account for electron-hole interactions \citep{chapter1-67,chapter1-68}.

\begin{figure}[H]
\begin{centering}
\includegraphics[width=1\textwidth,height=1\textheight,keepaspectratio]{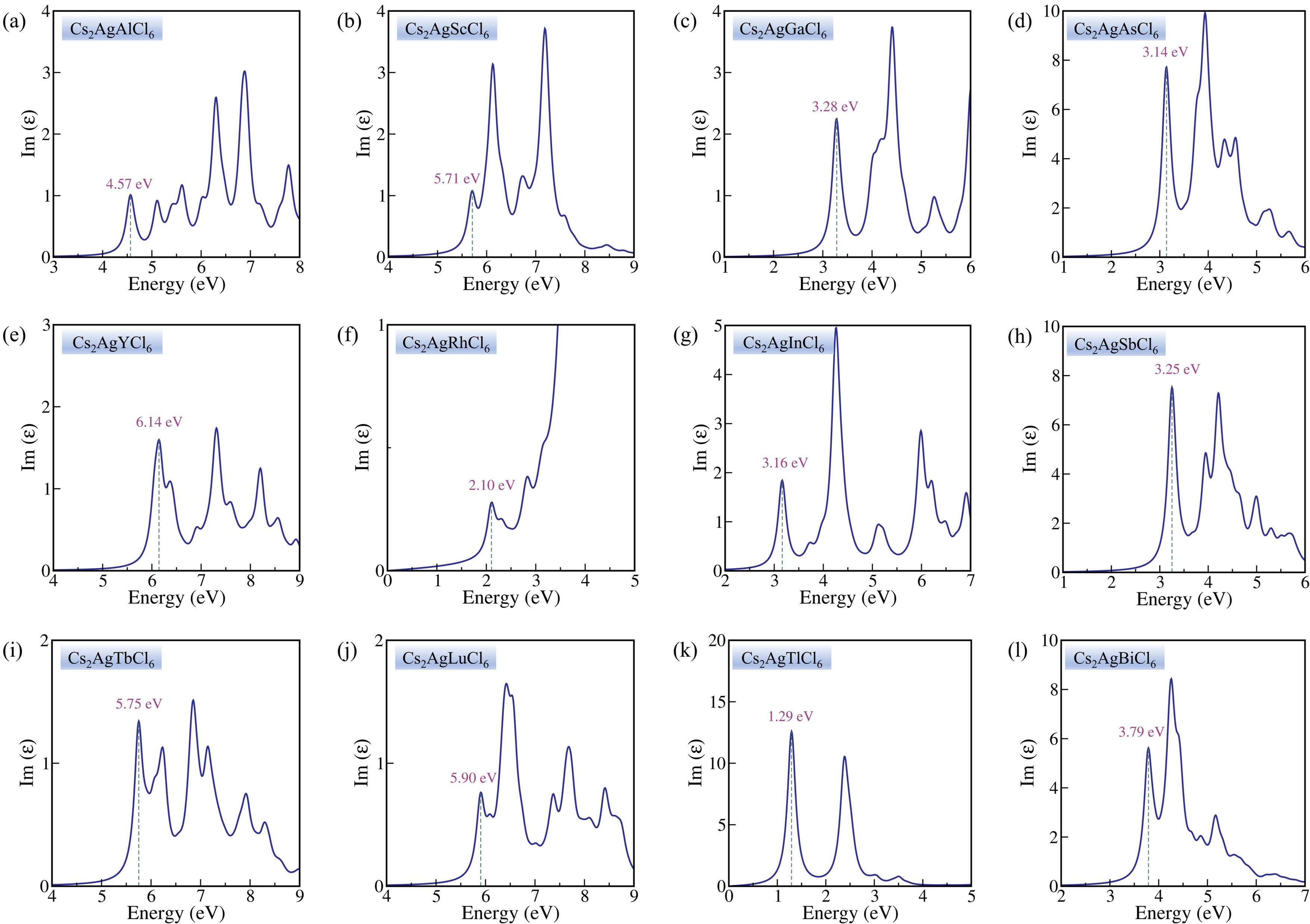}
\par\end{centering}
\caption{\label{Figure:3}Imaginary part {[}Im($\varepsilon$){]} of the dielectric
function for Cs$_{2}$AgM(III)Cl$_{6}$ HDPs obtained using BSE@G$_{0}$W$_{0}$@PBE
method, where M(III) = Al, Sc, Ga, As, Y, Rh, In, Sb, Tb, Lu, Tl,
and Bi.}
\end{figure}

The optical absorption spectra of Cs$_{2}$AgM(III)Cl$_{6}$ HDPs,
calculated using the BSE@G$_{0}$W$_{0}$@PBE method, are displayed
in Figure \ref{Figure:3}. It is observed that these HDPs display
an absorption onset spanning from the near-infrared to the ultraviolet
(UV) region. In indirect-bandgap materials, the absorption onset occurs
at energies significantly higher than their electronic bandgap, while
in direct-bandgap materials, the absorption edge closely aligns with
the electronic bandgap. This contrast highlights a key distinction:
direct-bandgap materials exhibit the first optical transition directly
between the VBM and the CBM, a feature notably absent in their indirect-bandgap
counterparts. Moreover, the first peak position or the optical bandgap
($E_{o}$) of these compounds is found within the range of 1.29$\lyxmathsym{\textendash}$6.14
eV, suggesting that electron-hole interactions significantly impact
the optical absorption spectra. The optical absorption spectra highlight
their potential applications for a wide range of optoelectronic devices.

In addition, the electronic dielectric constants ($\varepsilon_{\infty}$,
commonly referred to as the dielectric function at zero photon energy)
are obtained from the real part of the dielectric function, which
serves as a vital parameter for optoelectronic applications. A high
dielectric constant reduces the recombination rate of charge carriers,
thereby improving the performance of optoelectronic devices \citep{chapter2-48}.
The computed $\varepsilon_{\infty}$ values for Cs$_{2}$AgM(III)Cl$_{6}$
HDPs, determined using the BSE method, are presented in Table \ref{Table:3}.
These values serve as the basis for calculating various excitonic
and polaronic parameters, which are discussed in detail in the following
two sections.

\subsection{Excitonic Properties:}

Exciton parameters, including exciton binding energy ($E_{B}$), exciton
radius ($r_{exc}$), and exciton lifetime ($\tau_{exc}$), play a
pivotal role in determining the performance of optoelectronic devices.
The exciton binding energy ($E_{B}$) refers to the amount of energy
needed to dissociate an exciton into its individual electron ($e$)
and hole ($h$) components. When the exciton binding energy is lower,
the separation of electrons and holes becomes more efficient, leading
to enhanced photoelectric conversion efficiency in optoelectronic
devices. Theoretically, $E_{B}$ is estimated using first-principles
BSE calculations as \citep{chapter2-38,chapter5-18}: $E_{B}=E_{g}^{dir}-E_{o}$,
where $E_{g}^{dir}$ is the direct bandgap obtained from G$_{0}$W$_{0}$@PBE
calculations, and $E_{o}$ is the position of the first strong absorption
peak calculated using BSE@G$_{0}$W$_{0}$@PBE method. Table \ref{Table:4}
shows that for these HDPs, the exciton binding energies estimated
using the BSE@G$_{0}$W$_{0}$@PBE method range from 0.17 to 0.60
eV. The higher $E_{B}$ values observed in these systems indicate
pronounced excitonic effects, which can be advantageous for optoelectronic
applications such as light-emitting diodes and photodetectors.

It is worth noting that when the exciton binding energy ($E_{B}$)
significantly exceeds the longitudinal optical phonon energy ($\hbar\omega_{LO}$),
the electronic contribution to dielectric screening dominates. In
such cases, the ionic contribution can be neglected, leaving $E_{B}$
effectively unchanged \citep{chapter1-65,chapter1-66}. As shown in
Tables \ref{Table:4} and \ref{Table:5}, $E_{B}\gg\hbar\omega_{LO}$
for these HDPs, indicating that ionic screening to the dielectric
function can be safely ignored. This has been further validated using
the hydrogenic Wannier-Mott (WM) model. According to this model, the
exciton binding energy ($E_{B}$) of a screened Coulomb-bound $e-h$
pair is determined as \citep{chapter2-38}:

\begin{equation}
\mathrm{E_{B}=\left(\frac{\mu_{\mathit{dir}}^{*}}{m_{0}\varepsilon_{eff}^{2}}\right)R_{\infty}},\label{Eq:6}
\end{equation}

where $\mu_{dir}^{*}$ represents the reduced mass of the HDPs at
their direct band edges (see Table S6), $m_{0}$ is the rest mass
of an electron, $\mathrm{\varepsilon_{eff}}$ is the effective dielectric
constant, and $R_{\infty}$ is the Rydberg constant (13.6 eV). To
evaluate $E_{B}$ for the HDPs under investigation, it is essential
to compute the effective dielectric constant ($\mathrm{\varepsilon_{eff}}$).
The value of $\mathrm{\varepsilon_{eff}}$ lies between the electronic
(optical) dielectric constant ($\varepsilon_{\infty}$) and the static
dielectric constant ($\varepsilon_{static}=\varepsilon_{\infty}+\varepsilon_{ion}$),
where $\varepsilon_{ion}$ is the ionic dielectric constant.\textcolor{blue}{{}
}The values of $\varepsilon_{\infty}$ for these compounds are obtained
using the BSE method, while $\varepsilon_{ion}$ is calculated via
the DFPT method \citep{chapter1-60}. The corresponding dielectric
constants for the HDPs are listed in Table \ref{Table:3}. Therefore,
the upper ($E_{Bu}$) and lower ($E_{Bl}$) bounds of the exciton
binding energy are determined using the contributions from the optical
($\varepsilon_{\infty}$) and static ($\varepsilon_{static}$) dielectric
constants via Eq. \ref{Eq:6} and the results are presented in Table
S7. From these results, it is also observed that the upper bounds
(0.13$-$2.07 eV) for most HDPs closely match the $E_{B}$ values
obtained using the BSE@G$_{0}$W$_{0}$@PBE method, while the lower
bounds are significantly smaller. This indicates that the ionic contribution
to the dielectric constant is negligible, and $\mathrm{\varepsilon_{eff}}\rightarrow\varepsilon_{\infty}$,
implying that at high frequencies, dielectric screening is primarily
governed by electronic effects.

\begin{table}[H]
\caption{\label{Table:4}Calculated exciton parameters of Cs$_{2}$AgM(III)Cl$_{6}$
HDPs, where M(III) = Al, Sc, Ga, As, Y, Rh, In, Sb, Tb, Lu, Tl, and
Bi.}

\centering{}%
\begin{tabular}{ccccccc}
\hline 
Compounds & $E_{g}^{dir}$ ( eV) & $E_{o}$ ( eV) & $E_{B}$ ( eV) & $T_{exc}$ (K) & $r_{exc}$ (nm) & $|\phi_{n}(0)|^{2}$ (10$^{27}$ m$^{-3}$)\tabularnewline
\hline 
Cs$_{2}$AgAlCl$_{6}$ & 4.88 & 4.57 & 0.31 & 3594 & 0.63 & 1.25\tabularnewline
Cs$_{2}$AgScCl$_{6}$ & 6.31 & 5.71 & 0.60 & 6957 & 0.12 & 173.40\tabularnewline
Cs$_{2}$AgGaCl$_{6}$ & 3.45 & 3.28 & 0.17 & 1971 & 0.93 & 0.39\tabularnewline
Cs$_{2}$AgAsCl$_{6}$ & 3.41 & 3.14 & 0.27 & 3130 & 0.70 & 0.91\tabularnewline
Cs$_{2}$AgYCl$_{6}$ & 6.39 & 6.14 & 0.25 & 2899 & 0.34 & 8.26\tabularnewline
Cs$_{2}$AgRhCl$_{6}$ & 2.31 & 2.10 & 0.21 & 2435 & 0.57 & 1.74\tabularnewline
Cs$_{2}$AgInCl$_{6}$ & 3.39 & 3.16 & 0.23 & 2667 & 0.83 & 0.55\tabularnewline
Cs$_{2}$AgSbCl$_{6}$ & 3.62 & 3.25 & 0.37 & 4290 & 1.41 & 0.11\tabularnewline
Cs$_{2}$AgTbCl$_{6}$ & 6.27 & 5.75 & 0.52 & 6029 & 0.36 & 6.82\tabularnewline
Cs$_{2}$AgLuCl$_{6}$ & 6.18 & 5.90 & 0.28 & 3246 & 0.40 & 5.08\tabularnewline
Cs$_{2}$AgTlCl$_{6}$ & 1.47 & 1.29 & 0.18 & 2087 & 0.73 & 0.83\tabularnewline
Cs$_{2}$AgBiCl$_{6}$ & 4.13 & 3.79 & 0.34 & 3942 & 1.36 & 0.13\tabularnewline
\hline 
\end{tabular}
\end{table}

Next, we computed several additional excitonic parameters, such as
excitonic temperature ($T_{exc}$), exciton radius ($r_{exc}$) and
the probability of the wavefunction ($|\phi_{n}(0)|^{2}$) for the
$e-h$ pair at zero separation, by utilizing the $\mu_{dir}^{*}$
and $\varepsilon_{\infty}$, and the values of these parameters are
listed in Table \ref{Table:4}. $T_{exc}$ denotes the highest temperature
at which an exciton remains stable, and the thermal energy required
to dissociate an exciton is $E_{B}=k_{B}T_{exc}$, with $k_{B}$ being
the Boltzmann constant. The exciton radius ($r_{exc}$) is calculated
as follows \citep{chapter2-38,chapter8-9}:

\begin{equation}
r_{exc}=\frac{m_{0}}{\mu_{dir}^{*}}\mathrm{\varepsilon_{eff}}n^{2}r_{Ry},
\end{equation}

where $n$ is the exciton energy level and $r_{Ry}$ is the Bohr radius
(0.0529 nm). In our study, $\varepsilon_{\infty}$ is considered as
the $\varepsilon_{\mathrm{eff}}$ and $n=1$, which provides the smallest
exciton radius.

To calculate the exciton lifetime, we first calculated the probability
of the wave function ($|\phi_{n}(0)|^{2}$) for $e-h$ pairs at zero
separation using the following formula \citep{chapter2-38,chapter8-9}:

\begin{equation}
|\phi_{n}(0)|^{2}=\frac{1}{\pi(r_{exc})^{3}n^{3}}.
\end{equation}

The inverse of $|\phi_{n}(0)|^{2}$ can serve as a qualitative indicator
of the exciton lifetime ($\tau_{exc}$; for details, see the Supplemental
Material). A longer exciton lifetime corresponds to a lower carrier
recombination rate, which in turn enhances both the quantum yield
and conversion efficiency. This suggests that most of the HDPs studied
demonstrate longer exciton lifetimes, making them ideal for optoelectronic
applications, with the exception of those incorporating M(III) elements
like Sc, Y, Tb, and Lu.

\subsection{Polaronic Properties:}

To gain a deeper understanding of the fundamental limits of charge
carrier mobility in these HDPs, it would be beneficial to make predictions
using first-principles calculations. Polar materials often undergo
polaron formation, which influences the movement of charge carriers.
In polar semiconductors, the primary scattering mechanism near room
temperature is governed by the interaction between charge carriers
and the macroscopic electric field created by longitudinal optical
(LO) phonons \citep{chapter5-16,chapter5-18}. As a result, when calculating
mobility through theoretical methods, it is essential to account for
the polaron state, which emerges from the strong coupling between
charge carriers and phonons, rather than focusing solely on the free
carrier state. Fr\"ohlich developed a Hamiltonian to model the interaction
between low-density, independent charge carriers and polar optical
phonons \citep{chapter2-51}. Fr\"ohlich proposed a dimensionless parameter
($\alpha$) to quantify the strength of this interaction \citep{chapter5-16,chapter8-10},

\begin{equation}
\alpha=\frac{1}{4\pi\varepsilon_{0}}\frac{1}{2}\left(\frac{1}{\varepsilon_{\infty}}-\frac{1}{\varepsilon_{static}}\right)\frac{e^{2}}{\hbar\omega_{LO}}\left(\frac{2m^{*}\omega_{LO}}{\hbar}\right)^{1/2}
\end{equation}

where $\varepsilon_{0}$ represents the permittivity of free space,
$m^{*}$ denotes the effective mass of the carriers, and $\omega_{LO}$
refers to the angular frequency of the characteristic phonon. For
systems with multiple phonon branches, the average LO frequency is
obtained by computing the spectral average of all infrared-active
optical phonon branches \citep{chapter2-22} (for further details,
see the Supplemental Material). These LO frequencies are calculated
using the DFPT method. The values of $\alpha$ for electrons and holes
in Cs$_{2}$AgM(III)Cl$_{6}$ HDPs, which range from 2.24 to 9.99,
are provided in Table \ref{Table:5}. This implies that the HDPs under
investigation show an intermediate to strong electron (hole)-phonon
coupling \citep{chapter2-20}, with the hole-phonon interaction being
notably stronger than the electron-phonon interaction.

Polaron formation describes the process where a charge carrier (either
an electron or a hole) interacts with the lattice around it, leading
to a local distortion in the material. This interaction can lower
the quasiparticle (QP) energies, resulting in both electrons and holes
losing energy during the formation of polarons. The polaron energy
($E_{p}$) can be determined as follows \citep{chapter5-16}:

\begin{equation}
E_{p}=(-\alpha-0.0123\alpha^{2})\hbar\omega_{LO}.
\end{equation}

The QP gap, which originates from the polaron energy of electrons
and holes (see Table \ref{Table:5}), is calculated and compared with
the $E_{B}$ values presented in Table \ref{Table:4}. From this comparison,
it is evident that the energy of charge-separated polaronic states
is lower than that of the bound exciton states for the examined HDPs.
This implies that bound excitons (where electrons and holes stay closely
paired) are typically more stable than charge-separated polaronic
states (where electrons and holes are further apart) in these HDPs.

\begin{table}[H]
\caption{\label{Table:5}Calculated polaron parameters corresponding to electrons
($e$) and holes ($h$) in Cs$_{2}$AgM(III)Cl$_{6}$ HDPs, where
M(III) = Al, Sc, Ga, As, Y, Rh, In, Sb, Tb, Lu, Tl, and Bi.}

\centering{}%
\begin{tabular}{ccccccccccccc}
\hline 
\multirow{2}{*}{Compounds} & \multirow{2}{*}{$\omega_{LO}$ (THz)} & \multicolumn{2}{c}{$\alpha$} &  & \multicolumn{2}{c}{$E_{p}$ (meV)} &  & \multicolumn{2}{c}{$m_{p}/m^{*}$} &  & \multicolumn{2}{c}{$\mu_{p}$ (cm$^{2}$V$^{-1}$s$^{-1}$)}\tabularnewline
\cline{3-4} \cline{4-4} \cline{6-7} \cline{7-7} \cline{9-10} \cline{10-10} \cline{12-13} \cline{13-13} 
 &  & $e$ & $h$ &  & $e$ & $h$ &  & $e$ & $h$ &  & $e$ & $h$\tabularnewline
\hline 
Cs$_{2}$AgAlCl$_{6}$ & 4.57 & 4.04 & 7.00 &  & 80.26 & 143.88 &  & 2.08 & 3.39 &  & 15.17 & 1.41\tabularnewline
Cs$_{2}$AgScCl$_{6}$ & 4.79 & 5.18 & 7.57 &  & 109.30 & 164.14 &  & 2.53 & 3.70 &  & 4.92 & 0.82\tabularnewline
Cs$_{2}$AgGaCl$_{6}$ & 3.69 & 3.63 & 5.26 &  & 57.95 & 85.58 &  & 1.94 & 2.57 &  & 24.97 & 5.97\tabularnewline
Cs$_{2}$AgAsCl$_{6}$ & 3.61 & 3.31 & 4.28 &  & 51.50 & 67.35 &  & 1.83 & 2.17 &  & 22.48 & 8.69\tabularnewline
Cs$_{2}$AgYCl$_{6}$ & 3.91 & 5.00 & 8.54 &  & 85.94 & 152.81 &  & 2.46 & 4.25 &  & 8.77 & 0.74\tabularnewline
Cs$_{2}$AgRhCl$_{6}$ & 4.77 & 2.24 & 5.23 &  & 45.47 & 109.96 &  & 1.50 & 2.56 &  & 27.06 & 1.13\tabularnewline
Cs$_{2}$AgInCl$_{6}$ & 3.78 & 3.51 & 5.63 &  & 57.32 & 94.23 &  & 1.89 & 2.73 &  & 25.25 & 4.00\tabularnewline
Cs$_{2}$AgSbCl$_{6}$ & 3.35 & 3.69 & 4.67 &  & 53.52 & 68.51 &  & 1.96 & 2.32 &  & 18.99 & 7.86\tabularnewline
Cs$_{2}$AgTbCl$_{6}$ & 3.33 & 5.85 & 9.99 &  & 86.48 & 154.69 &  & 2.83 & 5.16 &  & 7.57 & 0.56\tabularnewline
Cs$_{2}$AgLuCl$_{6}$ & 3.58 & 4.64 & 8.70 &  & 72.72 & 142.79 &  & 2.31 & 4.34 &  & 12.04 & 0.71\tabularnewline
Cs$_{2}$AgTlCl$_{6}$ & 3.18 & 2.79 & 6.47 &  & 38.00 & 91.99 &  & 1.66 & 3.13 &  & 29.03 & 1.18\tabularnewline
Cs$_{2}$AgBiCl$_{6}$ & 3.18 & 4.09 & 4.92 &  & 56.57 & 68.71 &  & 2.10 & 2.42 &  & 15.43 & 7.64\tabularnewline
\hline 
\end{tabular}
\end{table}

Feynman introduced a novel method to tackle the Fr\"ohlich Hamiltonian,
which characterizes the interaction between an electron and a set
of independent phonon excitations that behave harmonically, all within
the framework of quantum field theory \citep{chapter2-23}. As the
electron traverses the lattice, it interacts with the perturbation
it generates, which decays exponentially with time. Using Feynman's
approach, the effective mass of the polaron ($m_{p}$) for our systems
is calculated as follows (for a small $\alpha$) \citep{chapter2-23}:

\begin{equation}
m_{p}=m^{\ast}\Big(1+\frac{\alpha}{6}+\frac{\alpha^{2}}{40}+...\Big)
\end{equation}

Table \ref{Table:5} indicate that the carrier-phonon coupling results
in an increase in the polaron effective mass by 50$\lyxmathsym{\textendash}$416\%,
which confirms the presence of intermediate to strong carrier-lattice
interactions.

To further confirm the impact of the increasing polaron effective
mass, the polaron mobility is also estimated using the Hellwarth polaron
model as \citep{chapter2-22},

\begin{equation}
\mu_{p}=\frac{\left(3\sqrt{\pi}e\right)}{2\pi c\omega_{LO}m^{*}\alpha}\frac{\sinh(\beta/2)}{\beta^{5/2}}\frac{w^{3}}{v^{3}}\frac{1}{K(a,b)}
\end{equation}
where $e$ is the electronic charge, $\beta=hc\omega_{LO}/k_{B}T$,
$w$ and $v$ both are the temperature-dependent variational parameters,
and $K(a,b)$ is a function of $\beta$, $w$, and $v$ (for details,
see the Supplemental Material). Polaron mobility represents the ease
with which a polaron moves through a lattice, and it is affected by
both its effective mass and the strength of interactions between the
carrier and the lattice. As the effective mass of the polaron increases
due to significant electron-phonon coupling, the mobility typically
decreases, resulting in slower motion of the polaron within the material.
This trend is observed in the systems under study (see Table \ref{Table:5}),
which show higher polaron mobility for electrons (4.92$-$29.03 cm$^{2}$V$^{-1}$s$^{-1}$)
compared to holes (0.56$-$8.69 cm$^{2}$V$^{-1}$s$^{-1}$). Regardless,
the investigated materials demonstrate relatively modest charge carrier
mobilities for electrons, along with desirable stability and non-toxicity,
making them a promising class of materials for optoelectronic devices.

\section{Conclusions:}

In summary, the impact of trivalent metal cation transmutation on
the structural, optoelectronic, excitonic, and polaronic properties
of lead-free Cs$_{2}$AgM(III)Cl$_{6}$ HDPs has been thoroughly investigated
through first-principles DFT- and MBPT-based simulations. Our result
reveals that the investigated compounds maintain a stable cubic double-perovskite
structure (space group $Fm\bar{3}m$), featuring ideal tolerance and
octahedral factors. In addition, these materials show remarkable thermodynamical,
dynamical, and mechanical stability. The electronic band structures
of Cs$_{2}$AgM(III)Cl$_{6}$ HDPs have been predicted through GW-based
calculations, revealing a broad range of bandgaps from 1.47 to 6.20
eV. By solving the BSE, the optical properties of these HDPs are precisely
determined, demonstrating absorption in the near-infrared to ultraviolet
region of the spectrum. Furthermore, we uncover an insignificant ionic
contribution to the effective dielectric screening used to calculate
the excitonic parameters. These perovskite materials exhibit moderate
exciton binding energies in the range of 0.17 to 0.60\,eV and generally
possess longer exciton lifetimes, except in the cases where M(III)
= Sc, Y, Tb, or Lu. Fr\"ohlich's mesoscopic model also reveals an intermediate
to strong electron (or hole)-phonon interaction, leading to reduced
polaronic mobility. The mobility ranges from 4.92 to 29.03 cm$^{2}$V$^{-1}$s$^{-1}$
for electrons and 0.56 to 8.69 cm$^{2}$V$^{-1}$s$^{-1}$ for holes.
It is also observed that the charge-separated polaronic states are
less stable than the bound excitons. Overall, this study not only
highlights the potential of Cs$_{2}$AgM(III)Cl$_{6}$ HDPs as promising
materials for diverse optoelectronic applications but also offers
valuable insights for the discovery and forecasting of next-generation
light-harvesting materials.
\begin{acknowledgments}
S.A. would like to acknowledge the Council of Scientific and Industrial
Research (CSIR), Government of India {[}Grant No. 09/1128(11453)/2021-EMR-I{]}
for Senior Research Fellowship. The authors acknowledge the High Performance
Computing Cluster (HPCC) \textquoteleft Magus\textquoteright{} at
Shiv Nadar Institution of Eminence for providing computational resources
that have contributed to the research results reported within this
paper.
\end{acknowledgments}

\bibliographystyle{apsrev4-2}
\bibliography{refs}

\end{document}